\documentclass[11pt]{article}
\usepackage{amsmath,amssymb,latexsym}
\usepackage{graphicx}

\newcommand\rf[1]{(\ref{eq:#1})}
\newcommand\lab[1]{\label{eq:#1}}
\newcommand\nonu{\nonumber}
\newcommand\br{\begin{eqnarray}}
\newcommand\er{\end{eqnarray}}
\newcommand\be{\begin{equation}}
\newcommand\ee{\end{equation}}

\newcommand\lb{\lbrack}
\newcommand\rb{\rbrack}

\newcommand\llb{\left\lbrack}
\newcommand\rrb{\right\rbrack}

\renewcommand\({\left(}
\renewcommand\){\right)}
\renewcommand\v{\vert}                     
\newcommand\bv{\bigm\vert}               
\newcommand\bgv{\bigg\vert}              

\newcommand\bc{\begin{center}}
\newcommand\ec{\end{center}}

\newcommand\partder[2]{\frac{{\partial {#1}}}{{\partial {#2}}}}

\renewcommand\b{\beta}

\renewcommand\d{\delta}

\newcommand\vareps{\varepsilon}
\newcommand\g{\gamma}
\newcommand\G{\Gamma}

\newcommand\h{\frac{1}{2}}
\renewcommand\k{\kappa}
\renewcommand\l{\lambda}
\renewcommand\L{\Lambda}
\newcommand\m{\mu}
\newcommand\n{\nu}
\renewcommand\o{\over}

\newcommand\p{\phi}
\newcommand\vp{\varphi}
\renewcommand\P{\Phi}
\newcommand\pa{\partial}

\newcommand\pr{\prime}

\renewcommand\r{\rho}
\newcommand\s{\sigma}

\renewcommand\t{\tau}
\renewcommand\th{\theta}

\newcommand\cA{{\mathcal A}}

\newcommand\cF{{\mathcal F}}

\newcommand{\ct}[1]{\cite{#1}}
\newcommand{\bib}[1]{\bibitem{#1}}

\newcommand\PRL[3]{\textsl{Phys. Rev. Lett.} \textbf{#1} (#2) #3}
\newcommand\NPB[3]{\textsl{Nucl. Phys.} \textbf{B#1} (#2) #3}

\newcommand\PRD[3]{\textsl{Phys. Rev.} \textbf{D#1} (#2) #3}

\newcommand\PLB[3]{\textsl{Phys. Lett.} \textbf{#1B} (#2) #3}
\newcommand\CQG[3]{\textsl{Class. Quantum Grav.} \textbf{#1} (#2) #3}

\newcommand\PR[3]{\textsl{Phys. Reports} \textbf{#1} (#2) #3}

\newcommand\IJMPA[3]{\textsl{Int. J. Mod. Phys.} \textbf{A#1} (#2) #3}

\newcommand\MPLA[3]{\textsl{Mod. Phys. Lett.} \textbf{A#1} (#2) #3}

\begin{document}

\title{Generalized Gauge Field Approach\\ To Lightlike Branes}

\author{E.I. Guendelman and A. Kaganovich\\
\small\it Department of Physics, Ben-Gurion University of the Negev,\\[-1.mm]
\small\it P.O.Box 653, IL-84105 ~Beer-Sheva, Israel  \\[-1.mm]
\small\it guendel@bgumail.bgu.ac.il , alexk@bgumail.bgu.ac.il\\
{} \\
Emil Nissimov and Svetlana Pacheva\\
\small\it Institute for Nuclear Research and Nuclear Energy,\\[-1.mm]
\small\it Bulgarian Academy of Sciences, Sofia, Bulgaria  \\[-1.mm]
\small\it email: nissimov@inrne.bas.bg, svetlana@inrne.bas.bg}

\maketitle

\begin{abstract}
We propose a general action describing the dynamics of {\em lightlike (LL)} 
$p$-branes in any odd $(p+1)$ world-volume dimensions. Next, we consider 
self-consistent coupling of \textsl{LL}-membranes ($p\! =\! 2$) 
to $D\! =\! 4$ Einstein-Maxwell system plus a $D\! =\! 4$ three-index antisymmetric 
tensor gauge field. The \textsl{LL}-brane serves as a material and charge
source for gravity and electromagnetism and, furthermore, it produces a
dynamical space-varying cosmological constant. We present static 
spherically-symmetric solutions where the space-time consists of two regions 
with different black-hole-type geometries and different values for dynamically 
generated cosmological constant, separated by the \textsl{LL}-brane which 
``straddles'' their common event horizon.
\end{abstract}

\vspace{.1in}

\section{Introduction}

In the context of non-perturbative string theory there arise several types of 
higher-dimensional membranes ($p$-branes, $Dp$-branes) which play a crucial role 
in the description of string dualities, microscopic physics of black holes, gauge 
theory/gravity correspondence \ct{duality-AdS-CFT}, cosmological brane-world 
scenarios \ct{R-S}, model building in high-energy particle phenomenology 
\ct{brane-model-build}, \textit{etc}.

There is a distinct class of branes -- {\em lightlike branes}, which are of 
particular interest in general relativity. They describe impulsive lightlike 
signals arising in various cataclysmic astrophysical events \ct{barrabes-hogan}. 
Lightlike membranes are basic ingredients in the so called ``membrane paradigm'' 
theory \ct{membrane-paradigm} of black hole physics. Furthermore, in the
context of the so called thin-wall description of domain walls coupled to 
gravity \ct{Israel-66,Barrabes-Israel-Hooft} they are able to provide quite
effective treatment of many cosmological and astrophysical effects.

In refs.\ct{Israel-66,Barrabes-Israel-Hooft} lightlike branes in the context of
gravity and cosmology have been extensively studied from a phenomenological point
of view, \textsl{i.e.}, by introducing them without specifying the Lagrangian 
dynamics from which they may originate. On the other hand, in a series of 
recent papers \ct{will-brane-1,will-trap-napoli} we have developed a new 
field-theoretic 
approach for a systematic description of the dynamics of lightlike branes starting 
from concise {\em Weyl-conformally invariant} actions. The latter are 
related to, but bear significant qualitative differences from, the standard 
Nambu-Goto-type $p$-brane actions\footnote{In ref.\ct{barrabes-israel-05} brane actions 
in terms of their pertinent extrinsic geometry have been proposed which generically
describe non-lightlike branes, whereas the lightlike branes are treated as a limiting
case.}.

The main aim of the present papers is to show that there exists a general
class of (not necessarily Weyl-conformally invariant) consistent Lagrangian 
theories of lightlike branes, which is universal in the sense that all these
theories yield physically equivalent solutions of the equations of motion, 
especially when coupled to bulk gravity-matter systems (see Section 4 below).

Our approach is based on two basic ingredients:
\begin{itemize}
\item
Employing alternative non-Riemannian integration measure (volume-form) 
\ct{TMT-basic,TMT-recent} in the 
actions of generally-covariant (reparametrization-invariant) field theories 
instead of (or, more generally, on equal footing with) the standard Riemannian 
volume form. 
\item
Employing auxiliary world-volume gauge field with a Lagrangian being an
arbitrary function of the standard Maxwell Lagrangian term.
\end{itemize}

Before proceeding to the main exposition let us briefly recall the standard 
Polyakov-type formulation of the ordinary (bosonic) Nambu-Goto $p$-brane action:
\be
S = -{T\o 2}\int d^{p+1}\s\,\sqrt{-\g}\, \Bigl\lb 
\g^{ab}\pa_a X^\m \pa_b X^\n G_{\m\n}(X) - \L (p-1)\Bigr\rb \; .
\lab{stand-brane-action}
\ee
Here $\g_{ab}$ is the ordinary Riemannian metric on the $p+1$-dimensional
brane world-volume with $\g \equiv \det\v\v \g_{ab}\v\v$. The world-volume indices
$a,b=0,1,\ldots ,p$; 
~$G_{\m\n}$ denotes the Riemannian metric in the embedding space-time with 
space-time indices $\m,\n=0,1,\ldots ,D-1$.
$T$ is the given \textsl{ad hoc} constant brane tension; the constant
$\L$ can be absorbed by rescaling $T$. 
The equations of motion w.r.t. $\g^{ab}$ and $X^\m$ read:
\be
T_{ab} \equiv \( \pa_a X^\m \pa_b X^\n - 
\h \g_{ab} \g^{cd}\pa_c X^\m \pa_d X^\n \) G_{\m\n} + \g_{ab} {\L\o 2}(p-1)
= 0 \; ,
\lab{stand-brane-gamma-eqs}
\ee
\be
\pa_a \(\sqrt{-\g}\g^{ab}\pa_b X^\m\) +
\sqrt{-\g}\g^{ab}\pa_a X^\n \pa_b X^\l \G^\m_{\n\l} = 0  \; ,
\lab{stand-brane-X-eqs}
\ee
where:
\be
\G^\m_{\n\l}=\h G^{\m\k}\(\pa_\n G_{\k\l}+\pa_\l G_{\k\n}-\pa_\k G_{\n\l}\)
\lab{affine-conn}
\ee
is the Christoffel connection for the external metric. In particular, when $p \neq 1$
Eqs.\rf{stand-brane-gamma-eqs} imply:
\be
\L \g_{ab} = \pa_a X^\m \pa_b X^\n G_{\m\n} \; .
\lab{stand-brane-metric-eqs}
\ee

Let us note the following properties of standard Nambu-Goto $p$-branes
manifesting their crucial differences w.r.t. the lightlike
branes discussed below.  Eq.\rf{stand-brane-metric-eqs} tells us that: (i) the 
induced metric on the Nambu-Goto $p$-brane world-volume is {\em non-singular};
(ii) standard Nambu-Goto $p$-branes describe intrinsically {\em massive} modes.

\section{Lightlike Branes. Action and Equations of Motion}

Let us consider the following new kind of $p$-brane action involving modified
world-volume integration measure density $\P (\vp)$ 
and an auxiliary (Abelian) world-volume gauge field $A_a$:
\be
S = - \int d^{p+1}\s \,\P (\vp)
\Bigl\lb \h \g^{ab} \pa_a X^{\m} \pa_b X^{\n} G_{\m\n} - L\!\( F^2\)\Bigr\rb
\lab{LL-brane}
\ee
\be
{\P (\vp) \equiv \frac{1}{(p+1)!} \vareps_{i_1\ldots i_{p+1}}
\vareps^{a_1\ldots a_{p+1}} \pa_{a_1} \vp^{i_1}\ldots \pa_{a_{p+1}} \vp^{i_{p+1}}}
\lab{mod-measure-p}
\ee
\be
F^2 \equiv F_{ab}(A) F_{cd}(A) \g^{ac}\g^{bd}
\lab{F2-def}
\ee
Here $\g_{ab}$ denotes the intrinsic Riemannian metric on the brane
world-volume, $\g = \det\Vert\g_{ab}\Vert$,
$F_{ab} = \pa_a A_b - \pa_b A_a$ and $a,b=0,1,\ldots,p; i,j=1,\ldots,p+1$.
$L\!\( F^2\)$ is an arbitrary function of the Maxwell Lagrangian term.
As we will see below, consistency of dynamics requires 
$F^2 L^{\pr}\!\( F^2\) >0$ (the prime on $L$ indicating derivative w.r.t. its
argument).

Rewriting the action \rf{LL-brane} in the following equivalent form:
\be
S = - \int d^{p+1}\!\s \,\chi \sqrt{-\g}
\Bigl\lb \h \g^{ab} \pa_a X^{\m} \pa_b X^{\n} G_{\m\n}  - L\!\( F^2\)\Bigr\rb
\quad, \quad
\chi \equiv \frac{\P (\vp)}{\sqrt{-\g}}
\lab{LL-brane-chi}
\ee
we see that the composite field $\chi$ plays the role of a dynamical
(variable) brane tension. Let us note the following differences of
\rf{LL-brane} (or \rf{LL-brane-chi}) w.r.t. 
the standard Nambu-Goto $p$-branes (in the Polyakov-like formulation) 
\rf{stand-brane-action}:

\vspace{.1in}
\begin{itemize}
\item
New non-Riemannian integration measure density $\P (\vp)$ instead of 
the usual $\sqrt{-\g}$, and no ``cosmological-constant'' term ($(p-1)\sqrt{-\g}$).
\item
Variable brane tension $\chi \equiv \frac{\P (\vp)}{\sqrt{-\g}}$.
\item
Auxiliary world-sheet gauge field $A_a$ entering via arbitrary non-linear
Lagrangian.
\item
Possibility for natural couplings of auxiliary $A_a$ to external world-volume
(``color'' charge) currents $J^a$.
\item
The action \rf{LL-brane} describes {\em intrinsically light-like} $p$-branes 
for any even $p$, \textsl{i.e.}, any odd-dimensional world-volume, see below. 
\item
For the special choice $L\( F^2\) = \sqrt{F^2}$ the brane action \rf{LL-brane} 
becomes Weyl-conformally invariant for {\em any} $p$
\ct{will-brane-1,will-trap-napoli}. Also, let us note that 
there are {\em no quantum conformal anomalies} in odd $(p+1)$ dimensions!
\end{itemize}

\vspace{.1in}
Employing the short-hand notations \rf{F2-def} and:
\br
\(\pa_a X \pa_b X\) \equiv \pa_a X^\m \pa_b X^\n G_{\m\n}
\lab{short-hand}
\er
for the induced metric, 
the equations of motion w.r.t. measure-building auxiliary scalars $\vp^i$ 
and $\g^{ab}$ read, respectively:
\br
\h \g^{cd}\(\pa_c X \pa_d X\) - L\!\( F^2\) = M \;
\Bigl( = \mathrm{integration ~const}\Bigr) \; ,
\lab{phi-eqs} \\
\h\(\pa_a X \pa_b X\) + 2 L^\pr\!\( F^2\) F_{ac}\g^{cd} F_{db} = 0 \; ,
\lab{gamma-eqs}
\er
where the latter can be viewed as $p$-brane analogues of the string Virasoro
constraints.

Eqs.\rf{phi-eqs}--\rf{gamma-eqs} have the following profound consequences.
First, from \rf{gamma-eqs} we obtain:
\be
\det\Vert \(\pa_a X \pa_b X\) \Vert = \( -4 L^\pr\! (F^2)\)^{p+1}
\(-\det\Vert\g_{ab}\Vert\) \(\det\Vert iF_{ab}\Vert\)^2
\lab{det-gamma-eqs}
\ee
For $(p+1)=$even world-volume dimensions the r.h.s. of \rf{det-gamma-eqs} is
strictly positive (because of the Lorentzian signature of $\g_{ab}$), 
whereas the determinant of the induced metric in the l.h.s of \rf{det-gamma-eqs}
should be negative conforming with the Lorentzian signatures of both world-volume
and embedding space-time metrics. Therefore, we conclude that the brane action
\rf{LL-brane} does not describe a consistent dynamics for even world-volume 
dimensions.

Next, taking the trace in \rf{gamma-eqs} and comparing with \rf{phi-eqs} implies
the following crucial relation for the Lagrangian function $L\( F^2\)$: 
\be
L\!\( F^2\) - 2 F^2 L^\pr\!\( F^2\) + M = 0 \; .
\lab{L-eq}
\ee
Eq.\rf{L-eq} can be viewed in two ways:

(a) For $M=0$ \rf{L-eq} is identically satisfied if we choose 
$L\!\( F^2\) = \sqrt{F^2}$. This is precisely the case of {\em Weyl-conformally
invariant} brane action \rf{LL-brane} which was introduced and extensively 
studied in Refs.\ct{will-brane-1,will-trap-napoli}.

(b) For arbitrary (non-zero) constant $M$ and arbitrary function $L\!\( F^2\)$ 
Eq. \rf{L-eq} determines $F^2$ as certain function of $M$, \textsl{i.e.}
\be
F^2 = F^2 (M) = \mathrm{const}
\lab{F2-const}
\ee
This is the generic case which will be discussed in what follows.

The third and most important implication of Eqs.\rf{gamma-eqs} is as follows.
Since $F_{ab}$ is anti-symmetric $(p+1)\times (p+1)$ matrix, then
$F_{ab}$ is {\em not invertible} in any odd $(p+1)$ -- it has at
least one zero-eigenvalue vector-field $V^a$: $F_{ab}V^b = 0$.
Therefore, for any odd $(p+1)$ the induced metric $\(\pa_a X \pa_b X\)$
on the world-volume of the $p$-brane model \rf{LL-brane} is {\em singular}
(as opposed to the ordinary Nambu-Goto brane, see \rf{stand-brane-metric-eqs}):
\br
\(\pa_a X \pa_b X\) V^b = 0 \quad ,\quad \mathrm{i.e.}\;\;
\(\pa_V X \pa_V X\) = 0 \;\; ,\;\; \(\pa_{\perp} X \pa_V X\) = 0
\lab{LL-constraints}
\er
where $\pa_V \equiv V^a \pa_a$ and $\pa_{\perp}$ are derivatives along the
tangent vectors in the complement of $V^a$. In particular, for $(p+1)\! =\! 3$
we have $V^a \simeq \h\frac{\vareps^{abc}}{\sqrt{-\g}}F_{bc}$.

Thus, we arrive at the following important conclusion:
every point on the world-surface of the $p$-brane \rf{LL-brane}
(for odd $(p+1)$) moves with the speed of light in a time-evolution along the
zero-eigenvalue vector-field $V^a$ of $F_{ab}$. Therefore, we will name
\rf{LL-brane} (for odd $(p+1)$) by the acronym {\em LL-brane}
(Lightlike-brane) model.

Finally, we get the equations of motion w.r.t. auxiliary gauge field $A_a$:
\br
\pa_b \( F_{cd}\g^{ac}\g^{bd}\sqrt{-\g}\,\chi\) = 0  \; ,
\lab{A-eqs}
\er
where relation \rf{F2-const} has been taken into account, 
and the equations of motion w.r.t. $X^\m$ :
\br
\pa_a \(\chi\sqrt{-\g} \g^{ab}\pa_b X^\m\) +
\chi\sqrt{-\g} \g^{ab}\pa_a X^\n \pa_b X^\l \G^\m_{\n\l} = 0 \; .
\lab{X-eqs}
\er

\vspace{.1in}
\textbf{Remark}. In what follows we will use a natural ansatz for the 
world-volume electric field $F_{0i}=0$ implying that 
$(V^a)=(1,\underline{0})$, \textsl{i.e.}, $\pa_V = \pa_0 \equiv \pa_\t$. 

\section{Special Case $p=2$. Coupling to Space-Time Maxwell and Rank-3 
Antisymmetric Tensor Gauge Fields}

Henceforth we will explicitly consider the special case $p=2$ of \rf{LL-brane},
\textsl{i.e.}, the lightlike membrane model:
\be
S = - \int d^3\s \,\P (\vp)
\Bigl\lb \h \g^{ab} \pa_a X^{\m} \pa_b X^{\n} G_{\m\n}(X) - L\! \( F^2\)\Bigr\rb
\lab{LL-membrane-0}
\ee
\be
\P (\vp) \equiv \frac{1}{3!} \vareps_{ijk}
\vareps^{abc} \pa_a \vp^i \pa_b \vp^j \pa_c \vp^k \quad ,\quad 
a,b,c =0,1,2\; ,\; i,j,k=1,2,3  \; .
\lab{mod-measure-3}
\ee

Invariance under world-volume reparametrizations allows to introduce the
standard (synchronous) gauge-fixing conditions:
\be
\g^{0i} = 0 \;\; (i=1,2) \quad ,\quad \g^{00} = -1 \; .
\lab{gauge-fix}
\ee
The ansatz $F_{0i}=0$, together with the Biancchi identity 
$\vareps^{abc}\pa_a F_{bc} = 0$
and \rf{gauge-fix} when inserted in \rf{F2-const}, implies:
\br
F^2 = 2B^2 = \mathrm{const} 
\quad ;\quad B = \h \frac{\vareps^{ij}}{\sqrt{\g^{(2)}}} F_{ij}
\quad ,\quad \g^{(2)} \equiv \det\Vert \g_{ij}\Vert \; ;
\lab{F2-const-0} \\
\pa_0 \(\vareps^{ij} F_{ij}\) = 0 \;\; \longrightarrow \;\; 
\pa_0 \g^{(2)} = 0 \; .
\lab{F2-const-1}
\er
Then the gauge-fixed equations motion for $A_a$ \rf{A-eqs} drastically simplify:
\be
\pa_i \chi = 0  \; ,
\lab{chi-eqs}
\ee
where $\chi \equiv \frac{\P (\p)}{\sqrt{-\g}}$ (the dynamical brane tension
as in \rf{LL-brane-chi}).

Employing \rf{gauge-fix}, the remaining gauge-fixed equations of motion 
w.r.t. $\g^{ab}$ and $X^\m$ read
(recall $\(\pa_a X \pa_b X\) \equiv \pa_a X^\m \pa_b X^\n G_{\m\n}$):
\br
\(\pa_0 X \pa_0 X\) = 0 \quad ,\quad \(\pa_0 X \pa_i X\) = 0  \; ,
\lab{constr-0}
\er
\br
\(\pa_i X\pa_j X\) - 2 c_1(M) \g_{ij} = 0 \quad ,\;\;
c_1(M) \equiv F^2 L^{\pr}\! (F^2)\bgv_{F^2 = F^2(M)} = \mathrm{const} \, ,
\lab{constr-vir}
\er
\br
\Box^{(3)} X^\m + \chi \sqrt{\g^{(2)}} \( - \pa_0 X^\n \pa_0 X^\l +
\g^{kl} \pa_k X^\n \pa_l X^\l \) \G^{\m}_{\n\l} = 0  \; ,
\lab{X-eqs-3-fix}
\er
\br
\Box^{(3)} \equiv - \pa_0 \(\chi \sqrt{\g^{(2)}} \pa_0 \) +
\pa_i \(\chi \sqrt{\g^{(2)}} \g^{ij} \pa_j \)
\lab{box-3}
\er

\textbf{Remark.} 
Note that \rf{constr-vir} coincides with the {\em space-like} part of the
Nambu-Goto brane constraint Eqs.\rf{stand-brane-metric-eqs}, whereas 
\rf{constr-0} drastically differ from their Nambu-Goto counterparts.

\vspace{.1in}
\textbf{Remark.} Consistency according to \rf{constr-vir} requires that
the constant:
\be
c_1 (M) \equiv F^2 L^{\pr}\! (F^2)\bgv_{F^2 = F^2(M)} >0  \; ,
\lab{c-positive}
\ee
therefore, an admissible choice for $L\!\( F^2\)$ is:
\be
L\!\( F^2\) = + \frac{1}{4} F^2 \quad ,\quad {i.e.}\;\; F^2 = 4M
\;\; ,\;\; c_1 (M) = M
\lab{Maxwell-wrong}
\ee
by virtue of Eq.\rf{L-eq}. The term \rf{Maxwell-wrong} is a Maxwell Lagrangian 
with a {\em wrong} sign, however, this is not a contradiction since in the context of
{\em LL}-brane \rf{LL-brane} the world-volume gauge field $A_a$ is an
auxiliary non-dynamical field.

\vspace{.1in}
\textbf{Remark.} In the special case of Weyl-conformally invariant lightlike branes
\ct{will-brane-1,will-trap-napoli}, \textsl{i.e.}, for 
$L\!\( F^2\) = \sqrt{F^2}$ and $M=0$ Eqs.\rf{F2-const-1} and \rf{constr-vir} 
change accordingly to ~$\pa_0 \(B\,\sqrt{\g^{(2)}}\) = 0$ and 
$\(\pa_i X\pa_j X\) - \sqrt{2} B \g_{ij} = 0$ (the magnetic field $B$ is not
necessarily constant in this case).

\vspace{.1in}
We can extend straightforwardly the {\em LL}-brane model \rf{LL-membrane-0}
via couplings to 
external space-time electromagnetic field $\cA_\m$ and, furthermore, to
external space-time rank 3 gauge potential $\cA_{\m\n\l}$ (Kalb-Ramond-type
coupling):
\br
S = - \int d^3\s \,\P (\vp)
\Bigl\lb \h \g^{ab} \pa_a X^{\m} \pa_b X^{\n} G_{\m\n} - L\!\( F^2\)\Bigr\rb
\nonu \\
-\; q\int d^3\s \,\vareps^{abc} \cA_\m \pa_a X^\m F_{bc} \;
- \;\frac{\b}{3!} \int d^3\s \,\vareps^{abc} \pa_a X^\m \pa_b X^\n \pa_c X^\l \cA_{\m\n\l}
\lab{LL-brane+A+A3}
\er
The second Chern-Simmons-like term in \rf{LL-brane+A+A3} is a special case 
of a class of Chern-Simmons-like couplings of extended objects to external
electromagnetic fields proposed in ref.\ct{Aaron-Eduardo}.

Let us recall the physical significance of $\cA_{\m\n\l}$ \ct{Aurilia-Townsend}. 
In $D=4$ when adding kinetic term for $\cA_{\m\n\l}$ coupled to gravity
(see Eq.\rf{E-M-LL} below), its field-strength:
\br
\cF_{\k\l\m\n} = 4\pa_{[\k}\cA_{\l\m\n]} = \cF \sqrt{-G} \vareps_{\k\l\m\n}
\lab{F4}
\er
with a single independent component $\cF$ produces {\em dynamical (positive)
cosmological constant}:
\be
K = \frac{4}{3}\pi G_N \cF^2 \qquad \Bigl( G_N - \mathrm{Newton ~constant}\Bigr) \; .
\lab{dynamical-cc}
\ee

The constraints \rf{constr-0}--\rf{constr-vir} (the gauged-fixed equations of
motion w.r.t. $\g^{ab}$) remain unaltered for the action \rf{LL-brane+A+A3}.
Using the same gauge choice ({$\g^{0i}=0, \g^{00}=-1$}) and ansatz
for the world-volume gauge field-strength ({$F_{0i}(A) = 0$}),
the equations of motion w.r.t. $A_a$ now acquire the form
(recall $\chi \equiv \frac{\P (\vp)}{\sqrt{-\g}}$ -- the brane tension,
$\cF_{\m\n} (\cA) = \pa_\m \cA_\n - \pa_\n \cA_\m$):
\br
\pa_i X^\m \pa_j X^\n \cF_{\m\n}(\cA) = 0 \quad ,\quad
\pa_i \chi +  \frac{\sqrt{2}q}{c_2 (M)} \pa_0 X^\m \pa_i X^\n \cF_{\m\n}(\cA) = 0 \; ,
\lab{A-eqs-1}
\er
where:
\be
c_2 (M) \equiv 2\sqrt{F^2} L^{\pr}\! (F^2)\bgv_{F^2 = F^2(M)} =
\mathrm{const} \: .
\lab{c2-def}
\ee
In particular $c_2 (M) = \sqrt{M}$ for the wrong-sign Maxwell choice
\rf{Maxwell-wrong}.

Eqs.\rf{A-eqs-1} tell us
that consistency of charged \textsl{LL}-brane dynamics implies that the
external space-time Maxwell field must have zero magnetic component normal
to the brane, as well as that the projection of the external electric field along
the brane must be proportional to the gradient of the brane tension.
Finally, the $X^\m$ equations of motion for \rf{LL-brane+A+A3} read:
\br
\Box^{(3)} X^\m + \chi \sqrt{\g^{(2)}}\( - \pa_0 X^\n \pa_0 X^\l +
\g^{kl} \pa_k X^\n \pa_l X^\l \) \G^{\m}_{\n\l}
\nonu \\
- 2qB\sqrt{\g^{(2)}} \pa_0 X^\n \cF_{\l\n}\, G^{\l\m}
- \frac{\b}{3!} \vareps^{abc} \pa_a X^\k \pa_b X^\l \pa_c X^\n G^{\m\r}\,
\cF_{\r\k\l\n} = 0 \; ,
\lab{X-eqs-1}
\er
where $\cF_{\r\k\l\n}$ is given as in \rf{F4} and 
$B = \sqrt{\h F^2} = \mathrm{const}$ ($B = \sqrt{2M}$ for the wrong-sign 
Maxwell choice, recall \rf{F2-const-0}, \rf{Maxwell-wrong}). 

\section{Bulk Einstein-Maxwell System Coupled to Lightlike Brane}

Now let us consider the coupled Einstein-Maxwell-\textsl{LL}-brane 
system adding also a coupling to a rank 3 gauge potential:
\br
{S = \int\!\! d^4 x\,\sqrt{-G}\,\llb \frac{R(G)}{16\pi G_N}
- \frac{1}{4} \cF_{\m\n}\cF^{\m\n} - \frac{1}{4! 2 } \cF_{\k\l\m\n}\cF^{\k\l\m\n}\rrb
+ S_{\mathrm{LL-brane}}} \; .
\lab{E-M-LL}
\er
Here $\cF_{\m\n} = \pa_\m \cA_\n - \pa_\n \cA_\m$,
$\cF_{\k\l\m\n} = 4 \pa_{[\k} \cA_{\l\m\n]} = \cF \sqrt{-G} \vareps_{\k\l\m\n}$
as above, and the \textsl{LL}-brane action is the same as in 
\rf{LL-brane+A+A3}.

The equations of motion for the \textsl{LL}-brane subsystem are the same as
\rf{constr-0}--\rf{constr-vir}, \rf{A-eqs-1}--\rf{X-eqs-1}, whereas the
Einstein, Maxwell and 3-index gauge field equations read:
\br
R_{\m\n} - \h G_{\m\n} R =
8\pi G_N \( T^{(EM)}_{\m\n} + T^{(rank-3)}_{\m\n} + T^{(brane)}_{\m\n}\) \; ,
\lab{Einstein-eqs}  \\
\pa_\n \(\sqrt{-G}G^{\m\k}G^{\n\l} \cF_{\k\l}\) + 
q \int\!\! d^3 \s\,\d^{(4)}\Bigl(x-X(\s)\Bigr)
\vareps^{abc} F_{bc} \pa_a X^\m = 0 \; ,
\lab{Maxwell-eqs}  \\
\vareps^{\l\m\n\k} \pa_\k \cF + \b\, \int\! d^3\s\, \d^{(4)}(x - X(\s))
\vareps^{abc} \pa_a X^{\l} \pa_a X^{\m} \pa_a X^{\n} = 0 \; .
\lab{F4-eqs}
\er
where in the last equation we have used relation \rf{F4}. 
The energy-momentum tensors read:
\br
T^{(EM)}_{\m\n} = \cF_{\m\k}\cF_{\n\l} G^{\k\l} - G_{\m\n}\frac{1}{4}
\cF_{\r\k}\cF_{\s\l} G^{\r\s}G^{\k\l} \; ,
\lab{T-EM} \\
T^{(rank-3)}_{\m\n} = \frac{1}{3!}\llb \cF_{\m\k\l\r} {\cF_{\n}}^{\k\l\r} -
\frac{1}{8} G_{\m\n} \cF_{\k\l\r\s} \cF^{\k\l\r\s}\rrb = - \h \cF^2 G_{\m\n} \; ,
\lab{T-rank3} \\
T^{(brane)}_{\m\n} = - G_{\m\k}G_{\n\l}
\int\!\! d^3 \s\, \frac{\d^{(4)}\Bigl(x-X(\s)\Bigr)}{\sqrt{-G}}\,
\chi\,\sqrt{-\g} \g^{ab}\pa_a X^\k \pa_b X^\l  \; .
\lab{T-brane}
\er
(recall $\chi\equiv\frac{\P(\vp)}{\sqrt{-\g}}$ -- the variable brane tension
\rf{LL-brane-chi}).

We will be looking for static spherically symmetric solutions of the
equations of motion \rf{Einstein-eqs}--\rf{F4-eqs} of the bulk 
gravity-matter system coupled to a charged \textsl{LL}-brane \rf{E-M-LL} (we
will assume spherical topology for the \textsl{LL}-brane surface). Notice that:

(a) The \textsl{LL}-brane serves as material and charge source for the bulk
gravity and electromagnetism;

(b) The resulting space-time will consist of two regions (one ``interior''
and one ``exterior'') separated by the
hypersurface of the \textsl{LL}-brane world-volume which will impose
non-trivial {\em matching conditions} across itself for the parameters of the 
(static) spherically symmetric geometries in both space-time regions.

(c) Consistency of dynamics of the \textsl{LL}-brane -- it should yield the
same solutions for the \textsl{LL}-brane equations of motion both from the
point of view of the interior as well as the exterior bulk gravity-matter.  

\vspace{.1in}
The general form of spherically-symmetric gravitational background in $D=4$ reads:
\br
{(ds)^2 = - A(r,t)(dt)^2 + B(r,t)(dr)^2 + 
C(r,t) \lb (d\th)^2 + \sin^2 (\th)\,(d\p)^2\rb}
\lab{spherical-symm-metric}
\er
Concerning point (c) above consider the following ansatz:
\br
X^0 \equiv t = \t \;\; ,\;\;
X^1 \equiv r = r (\t,\s^1,\s^2) \;\; ,\;\;
X^2 \equiv \th = \s^1 \quad ,\quad X^3 \equiv \p = \s^2
\lab{so3-ansatz} \\
\g_{ij} = a (\t,\s^1,\s^2) \( (d\s^1)^2 + \sin^2(\s^1) (d\s^2)^2\)
\phantom{aaaaaaaaaaaaa}
\lab{gamma-ansatz}
\er
and substitute it into the \textsl{LL}-brane equations of motion. We get:
\begin{itemize}
\item
Equations for $r (\t,\s^1,\s^2)$ from the lightlike constraints \rf{constr-0}:
\br
{\frac{\pa r}{\pa \t} = \pm \sqrt{\frac{A}{B}}  \quad ,\quad
\frac{\pa r}{\pa \s^i} = 0}
\lab{r-eqs}
\er
\item
A strong restriction on the gravitational background itself due to the
Virasoro-like constraints \rf{constr-vir}. Namely,
because of \rf{F2-const-1} the conformal factor $a$ in \rf{gamma-ansatz} must be
$\t$-independent. From this and \rf{constr-vir} we deduce:
\br
\pa_0 \(\pa_i X\pa_j X\) = 0 \;\; \longrightarrow \;\;
{\frac{d C}{d\t} \equiv \(\partder{C}{t} \pm
\sqrt{\frac{A}{B}}\, \partder{C}{r}\)\bgv_{t=\t,\; r=r(\t)} = 0}
\lab{C-eq}
\er
Eq.\rf{C-eq} tells us that the (squared) sphere radius $R^2 \equiv C (r,t)$ must
remain constant along the \textsl{LL}-brane trajectory. For static backgrounds
$R^2 \equiv C(r)$ Eqs.\rf{C-eq},\rf{r-eqs} imply:
\br
{r(\t) = r_0 \;\; (= \mathrm{const})\;\; , \quad A(r_0) = 0}
\lab{ll-horizon}
\er
Eq.\rf{ll-horizon} is of primary importance as it shows that
{\em the \textsl{LL}-brane automatically positions itself on the event horizon}.
\item
The Virasoro-like constraints \rf{constr-vir}, taking into account
\rf{ll-horizon} and \rf{gamma-ansatz}, imply:
\be
a = \frac{C(r_0)}{2 c_1 (M)} = \mathrm{const}
\lab{a-eq}
\ee
\item
\textsl{LL}-brane equations of motion \rf{X-eqs-1} for  $X^0\equiv t$ and
$X^1\equiv r$ turn out to be proportional to each other and reduce to an 
equation for the variable brane tension $\chi$ \rf{LL-brane-chi}:
\br
a \pa_\t\chi + \chi\, a\;\frac{\partder{}{t}\sqrt{AB} \pm \pa_r A}{\sqrt{AB}}
\pm \chi \,\frac{\pa_r C}{\sqrt{AB}} 
\nonu \\
\mp \frac{\sqrt{2}q}{c_2 (M)}\,
\frac{C}{\sqrt{AB}}\,\cF_{0r} \pm \b \cF C \sqrt{AB} = 0
\lab{chi-eq}
\er
where $\cF$ is the independent component of the rank 4 field-strength \rf{F4}
and the constant $c_2 (M)$ is the same as in \rf{c2-def} (here again one sets 
at the end $t=\t,\; r=r(\t)$).
\end{itemize}

Thus, following the same procedure as in the case of Weyl-conformally invariant 
lightlike branes \ct{will-trap-napoli} we arrive at the following static 
spherically symmetric solutions for the system \rf{E-M-LL}.
The bulk space-time consists of two regions separated by the \textsl{LL}-brane
sitting on (``straddling'') a common horizon of the former:
\br
{(ds)^2 = - A_{(\mp)}(r)(dt)^2 + \frac{1}{A_{(\mp)}(r)}(dr)^2 +
r^2 \lb (d\th)^2 + \sin^2 (\th)\,(d\p)^2\rb} \; ,
\lab{2-regions}
\er
where the subscript $(-)$ refers to the region inside, whereas the subscript 
$(+)$ refers to the region outside the horizon at 
$r=r_0 \equiv r_{\mathrm{horizon}}$ \rf{ll-horizon}. The interior
region is a Schwarzschild-de-Sitter space-time:
\br
A(r)\equiv A_{(-)}(r) = 1 - K_{(-)} r^2 - \frac{2G_N m_{(-)}}{r}
\;\; ,\quad \mathrm{for}\;\; r < r_0 \; ,
\lab{schwarzschild-dS-inside}
\er
whereas the exterior region is  Reissner-Nordstr\"{o}m-de-Sitter space-time:
\br
{A(r)\equiv A_{(+)}(r) = 1 - K_{(+)} r^2 - \frac{2G_N m_{(+)}}{r} +
\frac{G_N Q^2}{r^2}\;\; , \quad \mathrm{for}\;\; r > r_0} \; ,
\lab{RN-dS-outside}
\er
with Reissner-Norstr\"{o}m (squared) charge given by:
\be
Q^2 = 8\pi {\bar{q}}^2 r_{0}^4  \quad ,\quad \bar{q} \equiv \frac{q}{c_2 (M)} \; .
\lab{Q2-RN}
\ee
The rank 3 tensor gauge potential together with its Kalb-Rammond-type
coupling to the \textsl{LL}-brane produce via Eq.\rf{F4-eqs} a dynamical 
space-varying cosmological constant which is different inside and outside the 
horizon:
\be
K_{(\pm)} = \frac{4}{3}\pi G_N \cF_{(\pm)}^2 \;\; \mathrm{for} \;\; r \geq r_{0} \;\;
(\, r \leq r_{0}\,) \quad , \quad \cF_{(+)} = \cF_{(-)} - \b \; .
\lab{cosmolog-const}
\ee
The Einstein Eqs.\rf{Einstein-eqs} and the $X^\m$-brane Eqs.\rf{X-eqs-1}
yield two matching conditions for the normal derivatives w.r.t. the horizon
of the space-time metric components:
\br
\(\pa_r A_{(+)} - \pa_r A_{(-)}\)\!\!\bv_{r=r_0} = - 16\pi G_N \chi \; ,
\nonu \\
\(\pa_r A_{(+)} - \pa_r A_{(-)}\)\!\!\bv_{r=r_0} =
- \frac{r_0 (2{\bar{q}}^2 + \b^2) \pa_r A_{(-)}\bv_{r=r_0}}{2\chi + \b r_0 \cF_{(-)}}
\; .
\lab{matching}
\er
with $\bar{q}$ as in \rf{Q2-RN}.
The matching conditions \rf{matching} plus relation \rf{a-eq} allow
all physical parameters of the solution, \textsl{i.e.}, two spherically symmetric
black hole space-time regions ``soldered'' along a common horizon
materialized by the \textsl{LL}-brane, as well as the value of the
integration constant $M$, to be expressed in terms
of 3 free parameters $(q,\b,\cF)$ where (cf. Eq.\rf{LL-brane+A+A3}):

(a) $q$ is the \textsl{LL}-brane surface electric charge density;

(b) $\b$ is the \textsl{LL}-brane (Kalb-Rammond-type) charge w.r.t. rank 3 
space-time gauge potential $\cA_{\l\m\n}$;

(c) $\cF_{(-)}$ is the vacuum expectation value of the 4-index field-strength 
$\cF_{\k\l\m\n}$ in the interior region. 

\vspace{.1in}
For the common horizon radius $r_0$, the conformal factor $a$ of the internal
brane metric \rf{gamma-ansatz}, the Schwarzschild and Reissner-Nordstr\"{o}m 
masses $m_{(\mp)}$ we obtain ($\bar{q}$ is the same as in \rf{Q2-RN}):
\br
r_0^2 = \Bigl\lb 4\pi G_N \Bigl(\cF_{(-)}^2  - \b\cF_{(-)} + 
{\bar{q}}^2 +\frac{\b^2}{2}\Bigr)\Bigr\rb^{-1}
\; ,
\lab{r-horizon} \\
a = \Bigl({\bar{q}}^2 +\frac{\b^2}{2}\Bigr)
\Bigl\lb 2\pi G_N \Bigl( - \b\cF_{(-)} + {\bar{q}}^2 +\frac{\b^2}{2}\Bigr)^2\Bigr\rb^{-1} \; ,
\lab{a-sol}
\er
\br
m_{(-)} = \frac{r_0\,\(\frac{2}{3}\cF_{(-)}^2  - \b\cF_{(-)} + {\bar{q}}^2 +\frac{\b^2}{2} \)}
{2G_N \(\cF_{(-)}^2  - \b\cF_{(-)} + {\bar{q}}^2 +\frac{\b^2}{2}\)} \; ,
\lab{schwarzschild-mass}
\er
\br
m_{(+)} = m_{(-)} + \frac{r_0\,
\( 2{\bar{q}}^2 + \frac{2}{3}\b\cF_{(-)} - \frac{1}{3}\b^2\)
}{2G_N \(\cF_{(-)}^2  - \b\cF_{(-)} + {\bar{q}}^2 +\frac{\b^2}{2}\)}
\lab{RN-mass}
\er
For the brane tension we get accordingly:
\be
\chi = \frac{r_0}{2}\( {\bar{q}}^2 + \frac{\b^2}{2} - 2\b\cF_{(-)}\)
\lab{chi}
\ee
Inserting \rf{r-horizon}--\rf{a-sol} into relation \rf{a-eq} fixes the value ot 
the integration constant $M$. In particular, for the wrong-sign Maxwell choice 
\rf{Maxwell-wrong} and $\cF_{(-)} = 0$ we get $M = 1/4$.

Using expressions \rf{r-horizon}--\rf{RN-mass} we find
for the slopes of the metric coefficients $A_{(\pm)}(r)$ at $r=r_0$:
\br
\pa_r A_{(+)}\!\!\bv_{r=r_0} = - \pa_r A_{(-)}\!\!\bv_{r=r_0} \;\; ,\;\;
\pa_r A_{(-)}\!\!\bv_{r=r_0} = 8\pi G_N \chi 
\lab{A-slope-tension}
\er
with $\chi$ as in \rf{chi}. The typical form of $A (r)$ is depicted in Fig.1 below.
As shown in refs.\ct{will-trap-napoli}, this form of $A (r)$ creates a
potential ``well'' in the vicinity of the \textsl{LL}-brane lying on the
common horizon which acts as a trap for test particles falling toward the
horizon.


\vspace{.1in}
\textbf{Acknowledgements.}
{\small Three of us (A.K., E.N. and S.P.) are sincerely grateful for hospitality and
support to Prof. Branko Dragovich and the organizers of the 
\textsl{Fourth Summer School on Modern Mathematical Physics} (Belgrade, Serbia, 
Sept. 2006). E.N. and S.P. are supported by European RTN network
{\em ``Forces-Universe''} (contract No.\textsl{MRTN-CT-2004-005104}).
They also received partial support from Bulgarian NSF grant \textsl{F-1412/04}.
Finally, all of us acknowledge support of our collaboration through the exchange
agreement between the Ben-Gurion University of the Negev (Beer-Sheva, Israel) and
the Bulgarian Academy of Sciences.}

\begin{figure}[h]
\includegraphics{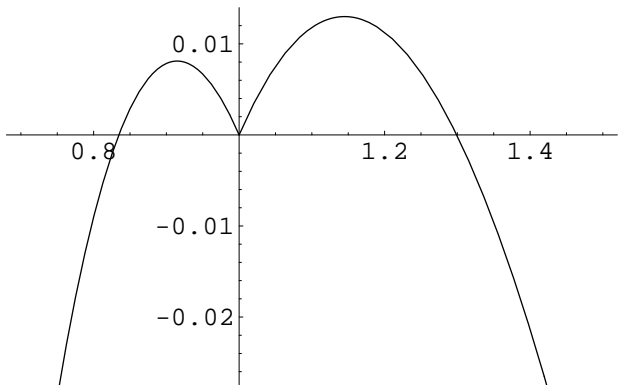}
\caption{Shape of $A (r)$ as a function of the dimensionless ratio 
$x \equiv r/r_0$}
\end{figure}


\end{document}